\documentclass[12pt,a4wide]{amsart}

\usepackage{amssymb}
 \usepackage{amsmath}
 \usepackage{hyperref}
\usepackage{graphicx}

\newtheorem{prop}{Proposition} 

\newtheorem{theorem}[prop]{Theorem}
\newtheorem{cor}[prop]{Corollary}

\newcommand{\pf}{\noindent{\em Proof. }}
\newcommand{\epf}{\hfill\hbox{\rule{6pt}{6pt}}}

\newcommand{\R}{{\mathbb R}}
\newcommand{\Z}{{\mathbb Z}}
\newcommand{\Rec}{\mathcal R}

\begin{document}

\title{Geometric medians in reconciliation spaces}

\author{K.~T.~Huber, V.~Moulton, M.-F. Sagot, B. Sinaimeri}
\address{KTH and VM: School of Computing Sciences,
University of East Anglia,  Norwich,
NR4 7TJ, UK. MFS and BS: Inria Grenoble - Rh\^{o}ne-Alpes; 
Inovall\'{e}e 655, avenue de l'Europe,
	Montbonnot, 38334 Saint Ismier cedex, France,
	Universit\'{e} de Lyon, F-69000, Lyon; Universit\'{e} Lyon 1;
 CNRS, UMR5558; 43
	Boulevard du 11 Novembre 1918, 69622 Villeurbanne cedex, France.
}
\date{\today}

\maketitle

\begin{abstract}
In evolutionary biology, it is common to study how 
various entities evolve together, for example, 
how parasites coevolve with their host, or genes with their species. 
Coevolution is commonly modelled by considering 
certain maps or {\em reconciliations} from one
evolutionary tree $P$ to another $H$, all of which 
induce the same map $\phi$ between the leaf-sets of $P$ and $H$
(corresponding to present-day associations).
Recently, there has been much interest
in studying spaces of reconciliations, which arise by 
defining some metric $d$ on the set  $\Rec(P,H,\phi)$
of all possible reconciliations between $P$ and $H$.\\
In this  paper, we study the following question:
How do we compute a {\em  geometric median} for 
a given subset $\Psi$ of $\Rec(P,H,\phi)$
relative to $d$, i.e. an  element $\psi_{med} \in \Rec(P,H,\phi)$ such that 
$$
\sum_{\psi' \in \Psi} d(\psi_{med},\psi') \le \sum_{\psi' \in \Psi} d(\psi,\psi')
$$
holds for all $\psi \in \Rec(P,H,\phi)$?
For a model where so-called host-switches or transfers are not allowed, and 
for a commonly used metric $d$ called the {\em edit-distance},
we show that although the cardinality of $\Rec(P,H,\phi)$ can be super-exponential, 
it is still possible to compute a geometric median 
for a set $\Psi$ in $\Rec(P,H,\phi)$ in polynomial time. We expect 
that this result could be useful for computing a summary or consensus  for
a set of reconciliations (e.g. for a set of suboptimal reconciliations).
\end{abstract}

\noindent {\bf Keywords:}
Reconciliation, Geometric median, 
Reconciliation space, Edit-distance,  Median \\

\noindent {\bf MSC[2008]:} 54E35, 05C05, 05C85, 92B05

\section{Introduction}
\label{section:introduction}
In phylogenetics, the reconciliation problem involves
trying to find a map that reconciles one leaf-labelled evolutionary tree
with another \cite{P94}.
It has important applications in areas such as ecology and genomics,
and arises in various situations.
For example, biologists are interested in understanding how 
parasite and host species \cite{D15}, genes and species \cite{D11},  
or species and habitats coevolve \cite{R78} (in what follows we
shall use terminology for host-parasite relationships to 
keep things concrete). 

\begin{figure}[t]
	\centering
	\includegraphics[scale=0.45]{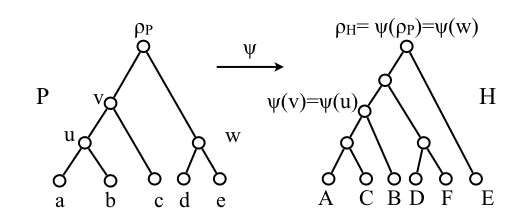}
	\caption{An example of a reconciliation map. Note 
		that $\phi$ is given by $\phi(a)=A,\ldots,\phi(e)=E$.} 
	\label{fig:rec}
\end{figure}

More formally, a {\em phylogenetic tree $T$} is a rooted, binary tree 
(i.e. every
vertex of $T$ that is not the root or a leaf has indegree  
1 and outdegree 2), which has  root vertex 
$\rho_T$ (with indegree 0 and outdegree 2).
Given a {\em host-parasite triple} $(P,H,\phi)$, that is,  
two phylogenetic trees $P$ and $H$ 
(the parasite and the host tree,  respectively), whose 
leaf-sets represent 
present-day species, and a map $\phi: L(P) \to L(H)$ (describing 
which parasite is currently on which host), a {\em reconciliation map}
is a map $\psi: V(P) \to V(H)$ 
which satisfies:
\begin{itemize}
	\item[(i)] The map $\psi$ restricted to the 
	leaf-set of $P$ is equal to $\phi$.
	\item[(ii)] If $v$ is a vertex in the interior of $P$, 
	then $\psi(v)$ is either strictly above or equal to
	$\psi(v')$, for any child $v'$ of $v$.
\end{itemize}
We present an example of such a map in Figure~\ref{fig:rec}.
Note that various definitions have been 
proposed for reconciliation maps (see 
e.g. \cite{D11}). These model evolutionary processes including  
cospeciation (a host and parasite speciate together), 
duplication (a parasite speciates on a host), loss 
(a host speciates but not its parasite) and  host-switches
(e.g. a parasite switches to another host).
In this paper, we are using the definition for a reconciliation map presented 
in \cite{D15,T11}, 
with the added assumption that we do not allow host-switches. 

In general, several algorithms have been developed to compute 
optimal and suboptimal reconciliations for a pair of trees 
relative to some predefined cost-function (cf. e.g. \cite{D11,D09}).
When host-switches are not allowed (as in this paper), 
collections of suboptimal reconciliations
can contain thousands of elements \cite{D09}, and for more
complex models (e.g. where host-switches are permitted), 
this can be the case even for collections of optimal reconciliations \cite{D15}.
It is thus quite natural to consider properties of 
the set of all possible reconciliations endowed with some metric
which also permits their comparison.  
These so-called \emph{reconciliation spaces} are of
growing importance in the literature \cite{B13,C15,D09,D12,W11}
and permit quantitative analysis of the behavior of reconciliation maps. 

In this paper, we are interested in the problem of
computing geometric medians in reconciliation spaces. 
In general, for $Y$ a finite set endowed with a metric $D$, and $Y' \subseteq Y$, 
an element $y^* \in Y$ is a  {\em geometric median} for $Y'$ in $Y$ if 
$$
\sum_{y'\in Y'} D(y^*,y') = \min\{  \sum_{y'\in Y'} D(y,y') \,: \, y \in Y \}.
$$
Such elements are useful as they can act as an element which
summarizes or forms a consensus for the set $Y'$. Within
computational biology, geometric medians (and the closely related concept of 
{\em centroids}) have been  
used in phylogenetics to form a consensus tree for a set of 
phylogenetic trees \cite{B01}, and in RNA secondary structure 
prediction to derive a consensus structure for a set of 
suboptimal RNA structures \cite{D05}. 
We therefore expect that being able to
compute geometric medians in reconciliation spaces 
should be a useful addition to the theory of reconciliations.

We now summarize the contents of the rest of the paper.
After presenting some preliminary definitions, 
in Section~\ref{sec:spaces}, 
we define the edit-distance, a metric on the set $\Rec(P,H,\phi)$
of all reconciliation maps for a host-parasite triple $(P,H,\phi)$.
Variants of this distance have been previously used to 
quantitatively analyse
collections of reconciliations (cf. e.g. \cite{D09}).
We then show that edit-distance
can be computed in a rather natural way
relative to the host tree. 
In  Section~\ref{sec:med}, we 
present some facts concerning medians, which 
we then use in Section~\ref{sec:med-rec} to define the
concept of a {\em median reconciliation} for a 
subset $\Psi$ of $\Rec(P,H,\phi)$.
In Section~\ref{sec:geom}, we then show that a 
median reconciliation is in fact a geometric median for 
$\Psi$ in $\Rec(P,H,\phi)$ relative to the edit-distance. We conclude in 
Section~\ref{sec:discuss}, with a brief discussion of some 
potential future directions.

\section{Preliminaries}\label{sec:prelim}

For a phylogenetic tree $T$, we denote the vertex set of $T$ 
by $V(T)$,  the set of interior vertices of $T$ by $V^o(T)= V(T) - L(T)$,
and the root by $\rho_T$. 
If $v \in V^o(T) $, we let $Ch(v)$ denote the set of children of $v$,
and if $v \in V(T) - \{\rho_T\}$, we let $par(v)$ denote the parent of $v$ in $T$.

We denote by $\succeq_T$ the partial order of $V(T)$ given by  $T$.
In case the context is clear, we just use $\succeq$. 
Also, we say
for vertices $x,y\in V(T)$ with $x\succeq y$ that $y$ is {\em below}
$x$ and that $x$ is {\em above} $y$. Furthermore, we say that  
$y$ is {\em strictly below} $x$ if $y$ is below $x$  
and $x\not=y$ and that  $x$ is {\em strictly above} $y$ 
if $x$ is above $y$  and $x\not=y$. In that case, we 
also put $x\succ y$. 
If $L$ is a subset of $L(T)$ of size at least two,
we let $lca_T(L) = lca(L)$ 
denote the {\em least common ancestor}
of the set $L$, that is, the lowest vertex in $T$ which is above 
every element of $L$ (with
respect to the ordering  $\succeq_T$). 
If $|L|=1$, then we 
set $lca_T(L)=x$ where $x$ is the unique element in $L$.

Now, let $(P,H,\phi)$ be a host-parasite triple.
For $v \in V(P)$, we let 
$$
m(v) = lca_H(\{ \phi(x) \,:\, x \in L(P)\mbox{ and } v\succeq_P x\}).
$$
We also let $A(v)$ be the subset of $V(H)$ given by
$$
A(v) = \{ v \in V(H) \,:\, \rho_H \succeq v \succeq m(v) \}.
$$

We now make some observations (cf. also \cite{D09}) --  we 
prove only (R2) as the rest are straight-forward to check:

\begin{itemize}
	\item [(R0)] If $v \in V^o(P)$ and $v' \in Ch(v)$, then $m(v) \succ m(v')$
	and $A(v) \subset A(v')$.  \epf
	
	\item [(R1)] If  $\psi \in \Rec(P,H,\phi)$, $x \in L(P)$, $v \in V(P)$ 
	and $v \succeq x$, then $\psi(v) \succeq \psi(x) = \phi(x)$.  \epf
	
	\item [(R2)] If $\psi\in \Rec(P,H,\phi)$, then 
	for all $v \in V(P)$ we have  $\psi(v) \in A(v)$.\\
	\pf If $v \in L(P)$ then the statement clearly holds. 
	Suppose now there exist some
	$v \in V^o(P)$, but $\psi(v) \not\in A(v)$. Since
	$m(v) \in A(v)$, it suffices to consider the following two cases:\\
	\noindent (i) $m(v) \succ \psi(v)$.  
	Note that by (R1),
	$\psi(v) \succeq \phi(x)$ for every 
	$x \in L(P)$ below $v$. Hence, 
	$\psi(v) \succ lca_H(\{ \phi(x) \,:\, x \in L(P)\mbox{ and } v\succeq_P x\})
	=m(v)$, which is impossible. 
	
	\noindent (ii) $m(v)$ and $\psi(v)$ are not comparable via $\succeq_H$.
	Then there  exists some $w\in A(v)-\{m(v)\}$ such that $w\succ \psi(v)$.
	Suppose $x \in L(P)$ is below $v$
	in $P$. By (R1), $\psi(v) \succeq \phi(x)$. But then $\phi(x)$ is not 
	below $m(v)$ in $H$. This 
	contradicts the definition of $m(v)$. \epf
	
	\item[(R3)] By 
	(R2), it follows that if 
	$\psi, \psi' \in \Rec(P,H,\phi)$, then for all 
	$v \in V(P)$, the vertices $\psi(v)$ and $\psi'(v)$ 
	are comparable in $H$ with respect
	to  the ordering $\succeq_H$. In particular, it also follows that 
	if $\{\psi_1, \dots, \psi_l\} \subseteq \Rec(P,H,\phi)$, 
	some $l\geq 1$, then for all $v \in V(P)$, 
	the ordering $\succeq_H$ induces a linear ordering on the set 
	$\{\psi_1(v), \dots, \psi_l(v)\}$.  \epf
\end{itemize}

\section{Reconciliation Spaces}\label{sec:spaces}

To compute a geometric median for some subset of $\Rec(P,H,\phi)$, we
first need to define a metric on $\Rec(P,H,\phi)$. In this paper, we 
focus on the 
{\em edit-distance}, $d_{edit}$, since edit-distances are  
commonly used to compare reconciliations (see e.g. \cite{D09}).

The edit-distance is defined as follows.
Given $\psi \in \Rec(P,H,\phi)$ and $w \in V(P)$ with 
$\psi(w) \not\in\{\rho_H,\psi(par(w))\}$, we define a map $\psi^{up}_w$ from $V(P)$ to $V(H)$ by 
setting $\psi^{up}_w(v)=par(\psi(v))$ if $v=w$
and $\psi^{up}_w(v) = \psi(v)$ if $v \in V(P)-\{w\}$. 
Moreover, given $\psi \in \Rec(P,H,\phi)$ and $w \in V^o(P)$ with 
$\psi(w) \succ m(w)$ and
$\psi(w) \neq \psi(v')$ for all $v' \in Ch(w)$, we define a 
map $\psi^{down}_w$ from $V(P)$ to $V(H)$ by 
setting $\psi^{down}_w(v)$ to 
be the (only) vertex in the set $A(w) \cap Ch(\psi(w))$ if $v=w$, 
and $\psi^{down}_w(v) = \psi(v)$ if $v \in V(P)-\{w\}$.
Now, given $\psi,\psi' \in \Rec(P,H,\phi)$, we define 
$d_{edit}(\psi,\psi')$
to be the smallest number of up/down operations required 
to change $\psi$ into $\psi'$. 
Note that this definition is closely related
to the edit-distance defined in \cite{D09}.

To prove our results concerning geometric  
medians, it is 
useful to have an alternative description of the edit-distance
which we now present.
If $v, w \in V(H)$, we let $d_H(v,w)$ be the length of the 
 (undirected) path in $H$ between $v$ and $w$.  Now, 
given $\psi,\psi' \in \Rec(P,H,\phi)$, we define the 
{\em path-distance} between 
$\psi$ and $\psi'$ by 
$$
d_{path}(\psi,\psi') = \sum_{v \in V(P)} d_H(\psi(v),\psi'(v)).
$$
It is easy to check that $d_{path}$ is a metric on  $\Rec(P,H,\phi)$ 
(i.e. \,$d_{path}(\psi,\psi')$ 
vanishes precisely when $\psi=\psi'$, 
it is symmetric
meaning $d_{path}(\psi,\psi')=d_{path}(\psi',\psi)$, and 
it also satisfies the triangle inequality
meaning $d_{path}(\psi,\psi'') \le d_{path}(\psi,\psi') + d_{path}(\psi',\psi'')$,
for all $\psi,\psi',\psi'' \in \Rec(P,H,\phi)$). We now 
prove the following theorem:

\begin{theorem}
	For all $\psi, \psi' \in \Rec(P,H,\phi)$,  $d_{edit}(\psi,\psi') = d_{path}(\psi,\psi')$.
	In particular, since  $d_{path}$ is a metric on $\Rec(P,H,\phi)$, so is $d_{edit}$.
\end{theorem}

Our proof for this theorem is very similar to the proof of \cite[Theorem 2]{D09}), but we include it 
for the sake of completeness; it immediately follows from the last 
of the following sequence of observations.\\

\noindent (Up) If $\psi \in \Rec(P,H,\phi)$ and $w \in V(P)$, and if
	$ \psi(w)\not\in\{\rho_H,\psi(par(w))\}$,
	then  $\psi^{up}_w \in \Rec(P,H,\phi)$.\\
	\pf This follows immediately, since if 
	$v\in V(P)-(\{par(w)\}\cup Ch(w))$ then $ \psi^{up}_w(v)=\psi(v)$.
	If $v \in Ch(w)$, then 
	$$
	\psi^{up}_w(w) = par(\psi(w)) \succ \psi(w) \succeq \psi(v) = \psi^{up}_w(v) 
	$$
	and if $v=par(w)$ then $\psi^{up}_w(v)=\psi(v)=
	\psi(par(w))\succeq par(\psi(w))=\psi^{up}_w(w)$.
	\epf\\	
	
\noindent (Down) If $\psi \in \Rec(P,H,\phi)$ and $w \in V(P)$ with 
	$\psi(w) \succ m(w)$ and 
	$\psi(w) \neq \psi(v')$ for all $v' \in Ch(w)$, 
	then  $\psi^{down}_w \in \Rec(P,H,\phi)$.\\
	\pf 
	Since $\psi(w) \not= m(w)$,
	it follows that 
	$\psi^{down}_w(w)\in A(w)$. 
	Moreover, since 
	$\psi(w) \neq \psi(v')$ for all $v' \in Ch(w)$
	and, by (R0), $A(v)\subset A(v')$ holds for all such 
	$v'$,
	we have 
	$\psi^{down}_w(w) \succeq \psi(v') = 
	\psi^{down}_w(v')$,
	for all $v'\in Ch(w)$. 
	Since $\psi^{down}_w(x)=\psi(x)=\phi(x)$ holds for all 
	$x\in L(P)$, 
	it follows that $\psi^{down}_w \in \Rec(P,H,\phi)$.
	\epf\\
	
	\noindent (E) Given $\psi,\psi' \in \Rec(P,H,\phi)$ distinct, 
	there exists a sequence 
	$(w_1,t_1)$, $(w_2,t_2),\dots,(w_p,t_p)$ with $w_i \in V(H)$ 
	and $t_i \in  \{up, down\}$, for 
	all $1 \le i \le p=d_{path}(\psi,\psi')$, 
	such that $\psi'$ is 
	the map obtained
	by successively applying up/down operations according to
	the pairs $(w_i,t_i)$, $1\leq i\leq p$,
	starting with the map $\psi$.  
	Moreover, no shorter sequence of operations exists for transforming 
	$\psi$ into $\psi'$. \\
	\pf 
	By the 
	assumption on  $\psi$ and $\psi'$  and (R3), 
	we may assume without loss of generality that 
	there exists some $w\in V(P)$ such that $\psi'(w) \succ \psi(w)$. Then either $\psi'(w)=\rho_H$ 
	or  we may assume without loss of generality 
	that $w$ is such 
	that, for all
	$w'\in V(P)$ strictly above 
	$w$, we have that $\psi(w')=\psi'(w')$.
	Hence, $\psi(w)\not=\psi(par(w))$.
	Starting with the map $\psi$, 
	it is 
	straightforward to check using (Up) that in either case
	we can apply a sequence of $d_H(\psi(w),\psi'(w))$ 
	operations of the form $(w,up)$ to obtain a new map 
	$\psi'' \in \Rec(P,H,\phi)$ with $\psi''(w) = \psi'(w)$ and
	$\psi''(v) = \psi(v)$ if $v \in V(P) -  \{w\}$. If there still exist
	vertices  $w'\in V(P)-\{w\}$  such that 
	$\psi'(w') \succ \psi(w')$, then
	we 	repeat this process until we obtain a map 
	$\psi^*\in \Rec(P,H,\phi)$ with the 
	property that $\psi^*(v) \succeq \psi'(v)$ holds
	for all $v \in V(P)$. 
	
	If $\psi^*=\psi'$, then Property (E) follows.
	Assume that
	$\psi^*\not=\psi'$. Then there must exist some $v\in V(P)$ such that
	$\psi^*(v)\succ \psi'(v)$.
	Out of all those $v \in V(P)$ with $\psi^*(v) \succ \psi'(v)$, choose 
	a vertex $w$ such that $d_P(w,\rho_P)$ is maximal. 
	We can then 
	transform $\psi^*$ into a new map in $\Rec(P,H,\phi)$
	by using a sequence of operations of the form 
	$(w,down)$. To see this, note first that 
	$\psi^*(w) \succ \psi'(w) \succeq m(w)$. Next, note  that there
	cannot exist some $v'\in Ch(w)$ such that
	$\psi^*(v')=\psi^*(w)$ as otherwise the choice
	of $w$ implies $\psi^*(w)=\psi^*(v')=\psi'(v')\preceq \psi'(w) \prec
	\psi^*(w)$ which
	is impossible. Since 
	$\psi^*\in \Rec(P,H,\phi)$, it follows by (Down) that 
	$(\psi^*)_w^{down}\in \Rec(P,H,\phi)$.   If we repeat 
	this process $d_H(\psi^*(w),\psi'(w))$ 
	times, 
	we eventually obtain a map that agrees with $\psi'$ on $w$ and 
	is equal to $\psi^*(v)$ for 
	all $v \in V(P)-\{w\}$. Repeating this process as many 
	times as necessary, we 
	eventually obtain the map $\psi'$.
	
	To obtain $\psi'$ from 
	$\psi$, we used  $d_{path}(\psi,\psi')$ operations.
	Moreover, we clearly need at least this number of operations.
	\epf

\section{Medians}\label{sec:med}

Before moving on to computing geometric medians for reconciliations, 
we first collect together some basic observations concerning medians. 

Given a multiset $A$ of real 
numbers, we let $med(A)$ denote 
the {\em median of $A$}. This is a real number, and is 
the ``middle" number of the set $A$ when 
the elements are arranged in order of magnitude. If the
cardinality of $A$ is even, the median is taken to be the real number that is 
half-way between the two middlemost numbers.

Given a real number $r$, we 
now
let $[r]$ denote the nearest 
integer to $r$ in case there
is only one, and to be the largest integer that is nearest to $r$ 
in case there are two nearest integers to $r$. 
For example, if $r=0.5$ then $[r]=\max\{0,1\}=1$, if  
$r=0.2$ then $[r]=0$, and if  $r=0.7$ then $[r]=1$.
Given a multiset $A$ of $m \ge 1$ 
integers, we define 
$zmed(A)$ to be $[med(A)]$. 
For example, if $A=\{1,1,2,3,4\}$ then $zmed(A)=2$, and if 
$A=\{1,1,2,3,4,5\}$ then $zmed(A)=3$.
Note that if $A = \{n_1,n_2,\dots,n_m\}$, then 
we also denote $med(A)$ and $zmed(A)$ by $med(n_1,n_2,\dots,n_m)$ and 
$zmed(n_1,n_2,\dots,n_m)$, respectively.
Also, if $m$ is odd, then $zmed(A)=med(A)$.

We now list some useful facts concerning the above definitions.

\begin{itemize}
\item[(M0)] Suppose that $A$ is a multiset of real numbers. 
	If $f:\R \to \R_{\ge 0}$ is
	the function given by 
	setting
	$$
	f(r) =  \sum_{a \in A} |a - r|
	$$
	for $r \in \R$, then $f(med(A)) \le f(r)$ for all $r \in \R$.\\
	\pf This is a well-known fact concerning medians. 
	Essentially it holds because,
	when $r$ moves away from $med(A)$, then $r$ moves away from at least as 
	many elements of $A$ as it approaches. Hence, $f$ attains its
	minimum over all $r \in \R$ at $med(A)$. \epf
		
	\item[(M1)] Suppose that $A, B$ are two multisets of integers 
	both containing $m \ge 1$ elements.
	Suppose that there exists an ordering $a_1,a_2,\dots,a_m$ of the 
	elements of $A$
	and an ordering $b_1, b_2,  \dots,  b_m$ of the elements of $B$ 
	such that $a_i \ge b_i$ for all $1 \le i \le m$.
	Then $med(A) \ge med(B)$ and $zmed(A) \ge zmed(B)$.\\
	\pf If $med(A) \ge med(B)$, then clearly $zmed(A) \ge zmed(B)$.
	
	To see that $med(A) \ge med(B)$, we consider the case where $m$ is odd; 
	the proof for $m$ even is similar.
	Let $a_{i_1}, a_{i_2}, \dots, a_{i_m}$ be an ordering of
	the elements of $A$ such that 
	$a_{i_1} \le a_{i_2} \le  \dots \le a_{i_m}$. 
	Then, $med(A) = a_{i_{\frac{m+1}{2}}}$ and,
	by assumption, at most $\frac{m+1}{2}-1$ elements in $B$ 
	(namely, $b_{i_{\frac{m+1}{2}+1}}, \dots, b_{i_m}$) can be greater than 
	$med(A)$, since if $1 \le j \le \frac{m+1}{2}$, 
	then $b_{i_j} \le a_{i_j} \le med(A)$. 
	Hence, $med(B) = b_{i_{\frac{m+1}{2}}}\le a_{i_{\frac{m+1}{2}}} =med(A)$.
	\epf \\
	
\item[(M2)] Suppose that $A$ is a multiset of integers, and $f$ is
	the function defined in (M0).
	Then $f(zmed(A))=f(med(A))$, and 
	so $f(zmed(A)) \le f(r)$ for all $r \in \R$.\\
	\pf If $A$ has an odd number of elements, we are done in view of (M0)
	since $zmed(A)=med(A)$.

	Suppose $A$ is even with cardinality $m$. If  
	$zmed(A) = med(A)$
	then we are done again in view of (M0). 
	Assume now that
	$zmed(A) \neq med(A)$. Then $zmed(A)$ is of the
	form $[r]$ where $r=med(A):=\frac{z'}{2}$ for some $z' \in \Z$. Therefore, 
	there exist two nearest integers $z_1, z_2$
	to $r$ that are both at distance $\frac{1}{2}$ from $r$. 
	Assume without loss of generality that  $z_1 > z_2$, so that 
	$z_1= r + \frac{1}{2}$, $z_2 = r -\frac{1}{2}$. 
	Then $z_1 = zmed(A)$. 
	But then for the function $f$ in (M0), we clearly have
	$f(r')=f(med(A))$ for all $r' \in [z_2,z_1]$.
	Statement (M2) now follows immediately.
	\epf
\end{itemize}	

\section{Median reconciliations}\label{sec:med-rec}

In this  section, we
define a special type of reconciliation $\psi_{med}= \psi^{\Psi}_{med}$
that can be associated to any subset $\Psi$ of $\Rec(P,H,\phi)$. 
In the next section, we 
prove that this 
is in actual fact a geometric median in the 
space $\Rec(P,H,\phi)$ endowed with the edit-distance.

Suppose $\Psi = \{\psi_1, \dots, \psi_l\} \subseteq \Rec(P,H,\phi)$, $l \ge 1$.
If $v \in V(P)$, then for $1 \le i \le l$, we let 
$$
n_i = d_H(m(v),\psi_i(v)).
$$
This is well defined by (R2) 
since $\psi_i(v)\in A(v)$, for all $1\leq i\leq l$.

We now define the map $\psi_{med}= \psi^{\Psi}_{med} $ from $V(P)$ to 
$V(H)$
by taking,  for $v \in V(P)$, $\psi_{med}(v)$ to be an element 
$w \in A(v) \subseteq V(H)$
such that $d_H(m(v),w) = zmed(n_1,n_2,\dots,n_l)$, for $v \in V(P)$.
Note that $\psi_{med}$ is well-defined
since $zmed(n_1,n_2,\dots,n_l)$ is an integer
and $zmed(n_1,n_2,\dots,n_l)\leq d_H(\rho_H,m(v))$. 
We now show that $\psi_{med}$ is a reconciliation.

\begin{theorem}
	$\psi_{med} \in \Rec(P,H,\phi)$.
\end{theorem}
\pf
First  note that  $\psi_{med}$ restricted to $L(P)$ is clearly equal to $\phi$.

Suppose now that $v \in V^o(P)$ and that 
$v' \in Ch(v)$. We need to show that
$\psi_{med}(v) \succeq \psi_{med}(v')$. 

First note that since $\psi_i(v)\in A(v)$ for all $1\leq i\leq l$,
Property (R0) implies
that $\{\psi_1(v),\dots, \psi_l(v), \psi_1(v'),\dots,\psi_l(v')\}$ 
is a subset of $A(v')$.
Moreover, $\psi_i(v) \succeq \psi_i(v')$ for all $1 \le i \le l$ as 
each $\psi_i$ is a reconciliation. 

Now, let $n_i = d_H(m(v),\psi_i(v))$ and $n'_i = d_H(m(v'),\psi_i(v'))$ 
for all $1 \le i \le l$.
Note that, by definition, $\psi_{med}(v)$ is equal to some 
$w\in A(v) \subset A(v')$ such that 
$d_H(m(v),w)=zmed(n_1,\dots,n_l)$,
and $\psi_{med}(v')$ is equal to some $w' \in A(v')$
such that 
$d_H(m(v'),w')=zmed(n_1',\dots,n_l')$.
For each $1 \le i \le l$, let 
\begin{eqnarray*}
p_i & = & n_i + d_H(m(v),m(v')) \\
& = & d_H(\psi_i(v),m(v)) + d_H(m(v),m(v'))\\
& = & d_H(\psi_i(v),m(v'))
\end{eqnarray*}
where the last equality holds in view of (R0).
Hence,
$d_H(m(v'),w) = zmed(p_1,\dots,p_l)$. 
Moreover, 
since $\psi_i(v) \succeq \psi_i(v') \succeq m(v')$ for all $1 \le i \le l$, 
it follows that $p_i \ge n'_i$.
By definition and (M1), it follows that 
$d_H(m(v'),w)=
zmed(p_1,\dots,p_l) \ge zmed(n'_1,\dots,n'_l)= d_H(m(v'), w')
$. Hence, $\psi_{med}(v) \succeq \psi_{med}(v')$, as required.
\epf\\

\noindent {\bf Remark:} Using similar arguments, we can also define a 
``minimum reconciliation'' for the set $\Psi$ as follows. 
Let $ \psi_{min}=\psi^\Psi_{min} :V(P) \to V(H)$ be given
by taking $\psi_{min}(v)$ to be a lowest element in  
$\{\psi_1(v), \dots, \psi_l(v)\}$ for $v \in V(P)$.
Note that $\psi_{min}$ is well-defined by (R3).
Moreover, $\psi_{min} \in \Rec(P,H,\phi)$: Indeed,  
$\psi_{min}$ restricted to $L(P)$ is clearly equal to $\phi$. Moreover,
if $v \in V^o(P)$, $v'  \in Ch(v)$, then for $i,j \in \{1,\dots,l\}$
such that $\psi_{min}(v) = \psi_i(v)$ and  $\psi_{min}(v') = \psi_j(v')$, we have 
$$
\psi_{min}(v') = \psi_j(v') \preceq \psi_i(v') \preceq \psi_i(v) = \psi_{min}(v).
$$
A similar  approach can be used to define  a``maximum reconciliation'' for $\Psi$.

\section{Geometric medians}\label{sec:geom}

In this section, we show that for a subset $\Psi$ of  $\Rec(P,H,\phi)$ endowed with the edit-distance, the 
reconciliation $\psi^{\Psi}_{med}$
is a geometric median for $\Psi$.
This will follow immediately 
from the following theorem:

\begin{theorem}\label{O1}
	Suppose that $T$ is a phylogenetic tree and that 
	$W = \{w_1,\dots,w_l\} \subseteq V(T)$, $l \ge 1$,
	is a subset of the set of vertices of some path $\gamma$ in $T$ 
	between $\rho_T$ and some vertex $s \in V(T)$.
	Let $q_i = d_T(w_i,s)$, $1 \le i \le l$, and let $u$  be a 
	vertex in $\gamma$ such that $d_T(u,s) = zmed(q_1,\dots,q_l)$. 
	Then for all $v' \in V(T)$,  
	\begin{equation}\label{inequal}
	\sum_{w \in W} d_T(v',w) \ge \sum_{w \in W} d_T(u,w).
	\end{equation} 
\end{theorem}
\pf Let $v' \in V(T)$.  First, suppose that $v'$ is a vertex in a path in $T$ between
$\rho_T$ and some leaf of $T$ that contains $\gamma$ as a subpath.

Let $A= \{q_1,\dots,q_l\}$, $\alpha = d_T(u,s)$ and $\beta=d_T(s,v')$ if 
$v'$ is above or equal to $s$ in $T$ and $\beta=-d_T(s,v')$ if $v$ is below $s$ in $T$.
Then, for the function $f$ in (M0), we have $f(\beta) \ge f(zmed(A)) $
in view of (M2). Hence,
$\sum_{i=1}^l |\beta-q_i| \ge \sum_{i=1}^l |\alpha - q_i|$, from 
which the theorem follows.
  
Suppose now that $v'$ is not of the above form. Then there
must exist some vertex $t$ in the path $\gamma$ such that $t \succ v'$. 
Using the same argument as above for $t$ instead of for $v'$, it follows that 
\begin{eqnarray*}
	\sum_{w \in W} d_T(v',w) & = & \sum_{w \in W}  (d_T(w,t) + d_T(t,v'))\\
	& =       & \sum_{w \in W}  d_T(w,t)  + |W| d_T(t,v')\\
	& \geq & \sum_{w \in W} d_T(u,w)   + |W| d_T(t,v')\\
	& \geq & \sum_{w \in W} d_T(u,w).
\end{eqnarray*}
\epf

\begin{cor}
	Suppose that $\Psi = \{\psi_1, \dots, \psi_l\} \subseteq \Rec(P,H,\phi)$, $l \ge 1$.
	Then $\psi^{\Psi}_{med}$ is a geometric 
	median for $\Psi$ in the space 
	$\Rec(P,H,\phi)$ endowed with the metric $d_{edit} (=d_{path})$.
\end{cor}
\pf Suppose that $\psi \in \Rec(P,H,\phi)$. Then by (R2), for $v \in V(P)$, taking 
$w_i = \psi_i(v)$, $u = \psi_{med}(v)$, $s = m(v)$ and $v'=\psi(v)$ in 
Theorem~\ref{O1}, we obtain
\begin{eqnarray*}
	\sum_{i=1}^m d_{path}(\psi_{med},\psi_i)
	& =   & \sum_{i=1}^l \sum_{v \in V(P)} d_H(\psi_{med}(v),\psi_i(v))\\
	& \le &  \sum_{i=1}^l \sum_{v \in V(P)} d_H(\psi(v),\psi_i(v))\\
	& =   & \sum_{i=1}^l d_{path}(\psi,\psi_i).
\end{eqnarray*}
\epf

Note that as a consequence of our results, 
we can compute a geometric 
median for a set $\Psi \subseteq  \Rec(P,H,\phi)$ in polynomial time.
Indeed, we can compute $d_H$ and the vertices $m(v)$, $v \in V(P)$
in polynomial time. Therefore, for each $v \in V(P)$, we can compute
the multiset of numbers $d_H(m(v),\psi(v))$, $\psi \in \Psi$,
the median of this multiset, 
and therefore $\psi_{med}(v)$, in polynomial time. 
It would be interesting to know if there is a more efficient way
to compute the map $\psi_{med}$.

\section{Discussion}\label{sec:discuss}

In this paper, we have 
described how to find a geometric median for 
a set of reconciliations within the space of 
all reconciliations endowed with the
path-distance (or, equivalently, the edit-distance). It would be of interest to understand 
properties of a geometric median. For example, reconciliations are
usually assigned some cost (see  
e.g. \cite{D09}), and it could be interesting to 
understand how the cost of the geometric median of a set
of reconciliations is related to the costs of each of the reconciliations in the set.
Also, we have focused on the edit-distance. However, it 
should be possible to define alternative metrics
on collections of reconciliations, and to potentially
derive geometric medians relative to these metrics. 

In another direction, as stated in the introduction, we 
considered one of the simplest models for reconciling trees. There are
more complex models which allow the inclusion of 
additional evolutionary processes (such as 
host-switches or, in the case of gene-species reconciliation, 
lateral gene transfer) \cite{T11}, and it would
be of interest to see whether geometric
medians can also be derived for these models.  
This could be useful since such models can
generate multiple optimal solutions \cite{D15}.
However, it could also be quite 
complicated as in our proofs we 
heavily relied on properties of
the median of a set of points in the real line, and for the 
more complex reconciliation models it is not clear
that such arguments can be applied.

Finally, in general the geometric median can be regarded
as a consensus for a set of  reconciliations.
It would be interesting to find other methods
for defining a consensus reconciliation and to understand 
how these are related to the geometric median
(e.g. we could try to define a centroid  reconciliation for a set which, 
roughly speaking, would correspond to the center of 
mass for the set).\\

\noindent{\bf Acknowledgement} All authors thank the Royal Society for its support.


\begin{thebibliography}{99}

\bibitem{B13}
M.\,Bansal, E.\,Alm, M.\,Kellis, 
Reconciliations revisited:
handling multiple optima when reconciling with duplication, transfer, and loss,
J. Comp. Bio. 
20(10) (2013) 738--754.

\bibitem{B01}
L.\,Billera, S.\,Holmes, K.\,Vogtmann,
Geometry of the space of phylogenetic trees,
Adv. in App. Math.
27 (2001) 733--767.

\bibitem{C15}
Y.\,Chan,  V.\,Ranwez, C.\,Scornavacca, 
Exploring the space of gene/species reconciliations with transfers,
J. Math. Biol.
71 (2015) 1179--1209.

\bibitem{C98}
M.\,A.\,Charleston, 
Jungles: a new solution to the host/parasite 
phylogeny reconciliation problem,
Math. Biosci.,  
149(2) (1998) 191--223.

\bibitem{D05}
Y.\,Ding, C.\,Chan, C.\,Lawrence,
RNA secondary structure prediction by centroids 
in a Boltzmann ensemble, 
RNA, 11 (2005) 1157--116.

\bibitem{D15}
B.\,Donati, C.\,Baudet, B.\,Sinaimeri, P.\,Crescenzi, M.\,F.\,Sagot,
{\sc{Eucalypt}}: Efficient tree reconciliation enumerator,
Alg Mol. Biol.  10(1) (2015) 3.

\bibitem{D11}
J.\,P.\,Doyon, V.\,Ranwez, V.\,Daubin, V.\,Berry,
Models, algorithms and programs for phylogeny reconciliation,
Brief. Bioinform. 12(5) (2011) 392--400.

\bibitem{D09}
J.\,Doyon, C.\,Chauve, S.\,Hamel,
Space of gene/species tree reconciliations and parsimonious models,
J. Comp. Biol. 16(10) (2009) 1399--1418.

\bibitem{D12}
J.\,Doyon, S.\,Hamel, C.\,Chauve, 
An efficient method for exploring the space of gene tree/species 
tree reconciliations in a probabilistic framework,
IEEE/ACM Trans. Comp. Bio. Bioinf. 9(1) (2012) 26--39.

\bibitem{H61}
C.\,A.\,R.\,Hoare, 
Algorithm 65: Find, 
Comm.\,ACM., 4(7) (1961) 321--322.

\bibitem{P94}
R.\,Page, 
Maps between trees and cladistic analysis of historical 
associations among genes, organisms, and areas,
Sys. Bio. 43(1) (1994) 58--77.

\bibitem{R78}
D.\,Rosen,
Vicariant patterns and historical explanation in biogeography,
Syst. Biol. 27(2) (1978) 159--88.

\bibitem{T11}
A.\,Tofigh, M.\,Hallett, J.\,Lagergren, 
Simultaneous identification of duplications and lateral gene transfers,
IEEE/ACM Trans. Comp. Bio. Bioinf. 8(2) (2011) 517--535.

\bibitem{W11}
T.\,Wu, L.\,Zhang,
Structural properties of the reconciliation 
space and their applications in enumerating nearly-optimal reconciliations 
between a gene tree and a species tree,
BMC Bioinf. (2011) 12(Suppl 9):S7.

\end{thebibliography}
\end{document}